\documentclass[fleqn,10pt]{wlscirep}
\usepackage[utf8]{inputenc}
\usepackage[T1]{fontenc}
\usepackage{xcolor}
\usepackage{layouts}
\usepackage{booktabs}
\title{A Non-negative Matrix Factorization Based Method for Quantifying Rhythms of Activity and Sleep and Chronotypes Using Mobile Phone Data}

\author[1,2,*]{Talayeh Aledavood}
\author[2]{Ilkka Kivim\"aki}
\author[3,4]{Sune Lehmann}
\author[2]{Jari Saram\"aki}
\affil[1]{School of Interactive Computing, Georgia Institute of Technology}
\affil[2]{Department of Computer Science, Aalto University, Espoo, Finland}
\affil[3]{Department of Applied Mathematics and Computer Science, Technical University of Denmark, Kongens Lyngby, Denmark}
\affil[4]{The Niels Bohr Institute, University of Copenhagen, Copenhagen, Denmark}
\affil[*]{talayeh@gatech.edu}

\begin{abstract}

Human activities follow daily, weekly, and seasonal rhythms. The emergence of these rhythms is related to physiology and natural cycles as well as social constructs. 
The human body and biological functions undergo near 24-hour rhythms (circadian rhythms). 
The frequency of these rhythms is more or less similar across people, but its phase is different. 
In the chronobiology literature, based on the propensity to sleep at different hours of the day, people are categorized into morning-type, evening-type, and intermediate-type groups called \textit{chronotypes}. 
This typology is typically based on carefully designed questionnaires or manually crafted features drawing on data on timings of people's activity. 
Here we develop a fully data-driven (unsupervised) method to decompose individual temporal activity patterns into components. 
This has the advantage of not including any predetermined assumptions about sleep and activity hours, but the results are fully context-dependent and determined by the most prominent features of the activity data. 
Using a year-long dataset from mobile phone screen usage logs of 400 people, we find four emergent temporal components:
morning activity, night activity, evening activity and activity at noon. 
Individual behavior can be reduced to weights on these four components. 
We do not observe any clear emergent categories of people based on the weights, but individuals are rather placed on a continuous spectrum according to the timings of their activities. 
High loads on morning and night components highly correlate with going to bed and waking up times. 
Our work points towards a data-driven way of categorizing people based on their full daily and weekly rhythms of activity and behavior, rather than focusing mainly on the timing of their sleeping periods.

\end{abstract}
\begin{document}

\flushbottom
\maketitle
%
%
\thispagestyle{empty}

\newpage

\section{Introduction}
Human lives are defined by rhythms of different frequency: daily, weekly, seasonal, and annual, among others. 
The most prominent rhythms in our lives are rooted in the day-night cycle~\cite{foster2014rhythms}. 
From body cell functions to social activities and interactions, many aspects of human lives follow diurnal rhythms~\cite{panda2002circadian}. 
Sleep-wake cycles are regulated by circadian rhythms which are internal body processes~\cite{edery2000circadian}.
Sleep is important to our health and well-being and enables us to restore physically and mentally to pursue our daily activities~\cite{irwin2015sleep}. 
Circadian rhythms have a similar near 24-hour length in all humans, but their phase varies from one person to another which results in different sleeping hours~\cite{kerkhof1985inter}. 
There is a large body of literature that shows that people with later-phase circadian rhythms are at risk of various physical and mental issues~\cite{fabbian2016chronotype, antypa2016chronotype, romo2020evening}. 
Therefore understanding and measuring these rhythms are crucial.  

In chronobiology, people are commonly divided into three categories referred to as ``chronotypes''~\cite{adan2012circadian} according to the difference in the phase of circadian rhythms and the propensity to sleep at different hours. 
The three widely-accepted chronotypes are morning-type, evening-type, and intermediate-type~\cite{levandovski2013chronotype}. Chronotypes are often measured with one of the various available questionnaires which have been developed for this purpose since the 1970s~\cite{levandovski2013chronotype}. 
While chronotype questionnaires are widely used, they come with some shortcomings. 
First, like any other questionnaire they rely on a person's description of their behavior rather than measuring the behavior \textit{in situ}. 
Also, most of them have cut-off boundaries, which might not be appropriate for the population under study. 
For example, the most well-known chronotype questionnaire, the Morningness-Eveningness Questionnaire (MEQ)~\cite{horne1976self}, was originally designed based on a cohort of adults with ages between 18 to 32 years old~\cite{levandovski2013chronotype}
\footnote{There are also questionnaires such as the Munich ChronoType Questionnaire (MCTQ)~\cite{roenneberg2003life} which are not based on pre-defined thresholds and assume that chronotype is a continuous variable rather than a categorical one~\cite{roenneberg2015human}.}.

In this work, instead of using questionnaires relying on people's memories, we measure their sleep and activity patterns based on digital traces that 
are recorded in an accurate and 
unobtrusive manner from mobile phones. 
Instead of matching our subjects' activity patterns to another cohort, we use Non-negative Matrix Factorization (NMF)~\cite{cichocki2009nonnegative} 
to automatically learn the dominant patterns in the data. 
We use this unsupervised method to reduce the dimensionality of the data and to extract four dominant rhythms within the population's activity patterns. 
This approach minimizes the a priori assumptions on the sleep and activity hours of the study participants and lets these patterns emerge from the data. 
Using NMF, we reduce each person's rhythm into weights on the extracted components. 
This way we can measure the typical timings of a person's sleep and activity without having the common biases of questionnaires or enforcing arbitrary cut-off thresholds to the data. 
We also do not require any prior knowledge about the population under study.

Our work is part of a rapidly growing field in which digital traces that people leave behind, such as data produced by mobile phones, are used as a proxy for human activity and to measure temporal patterns of their behavior \cite{cuttone2014inferring, aledavood2015daily,  aledavood2017temporal, urena2020going}. 
These traces allow us to measure people's behavioral patterns unobtrusively in the wild. 
Continued evolution of mobile phones and especially the ubiquity of smartphones has opened up possibilities to collect high-resolution data from individuals and study their behavior at a level of detail not previously possible~\cite{stopczynski2014measuring}. 
In the past years, multiple studies have gathered detailed data from individuals with the aim of studying and quantifying their behavior \cite{eagle2006reality, wang2014studentlife, stopczynski2014measuring, mattingly2019tesserae}. 
More recently, high-resolution data from phones have been used to study sleeping and resting patterns of people~\cite{monsivais2017seasonal, cuttone2017sensiblesleep, aledavood2018social, aledavood2019smartphone, martinez2020improved}. 
Our work introduces a new method to study these patterns that can be applied to data from any of these (and other similar) studies.

In this work, we use data from the Copenhagen Networks Study, which has gathered data from around 1000 university students for over two years~\cite{stopczynski2014measuring}. 
In this study, participants were given identical phones, and, with their consent, data on their phone usage and behavior were collected and shared with researchers. 
We used timestamps of events where the phone screen turns on (screen-on events) to measure dominant activity and sleep rhythms of the study participants. 
In this work, we also study in detail how the component weights of the decomposition of rhythms correlate with one another within the population. 
Furthermore, we investigate how these weights correlate with sleep and wake-time hours as well as the sleep duration for different people.

\section{Implementation and Results}
\begin{figure}[h!]
\centering
\includegraphics[width=\linewidth]{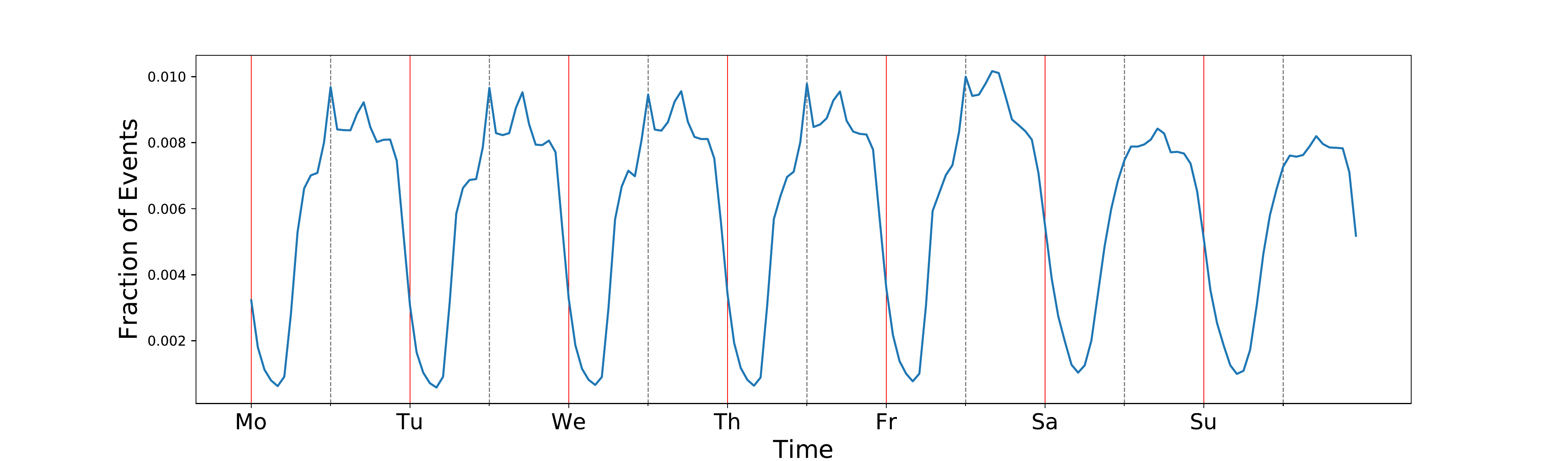}
\caption{The average normalized activity rhythm of everybody included in the analysis. The x-axis shows a week (hourly bins). The red lines mark the beginning of each day and the grey dashed lines mark 12:00 noon on each day.}
\label{fig:population-rhythm}
\end{figure}

Out of over 800 students who have provided data to the Copenhagen Networks Study during the year 2014, we included 400 in this analysis after filtering out individuals with low data quality (see section~\ref{sec:methods} for details on data pre-processing). 
We use screen-on events as a proxy of the times that the person is active throughout the day. 
While this measure does not capture all types of activity, it is, for example, a good measure of when a person awake and using the phone. 
By studying the temporal activity patterns of phone usage over a long period of time, we can form a picture of the behavioral patterns of the person. 
We aggregated the weekly data for each person over the course of one year into $7\times 24=168$ one-hour bins (Monday morning to Sunday night). 
For each person we normalize the sum of the time series to unity and derive the \textit{activity rhythm} of the person. 
The average of the activity rhythms of all the 400 individuals is shown in Fig.~\ref{fig:population-rhythm}. 
The average rhythm shows clear daily periodicity and a lower activity level during weekends, reflecting the common pattern in individual rhythms. 
The lowest level of activity coincides with night times. 

\begin{figure}[!ht]
\centering
\includegraphics[width=\linewidth]{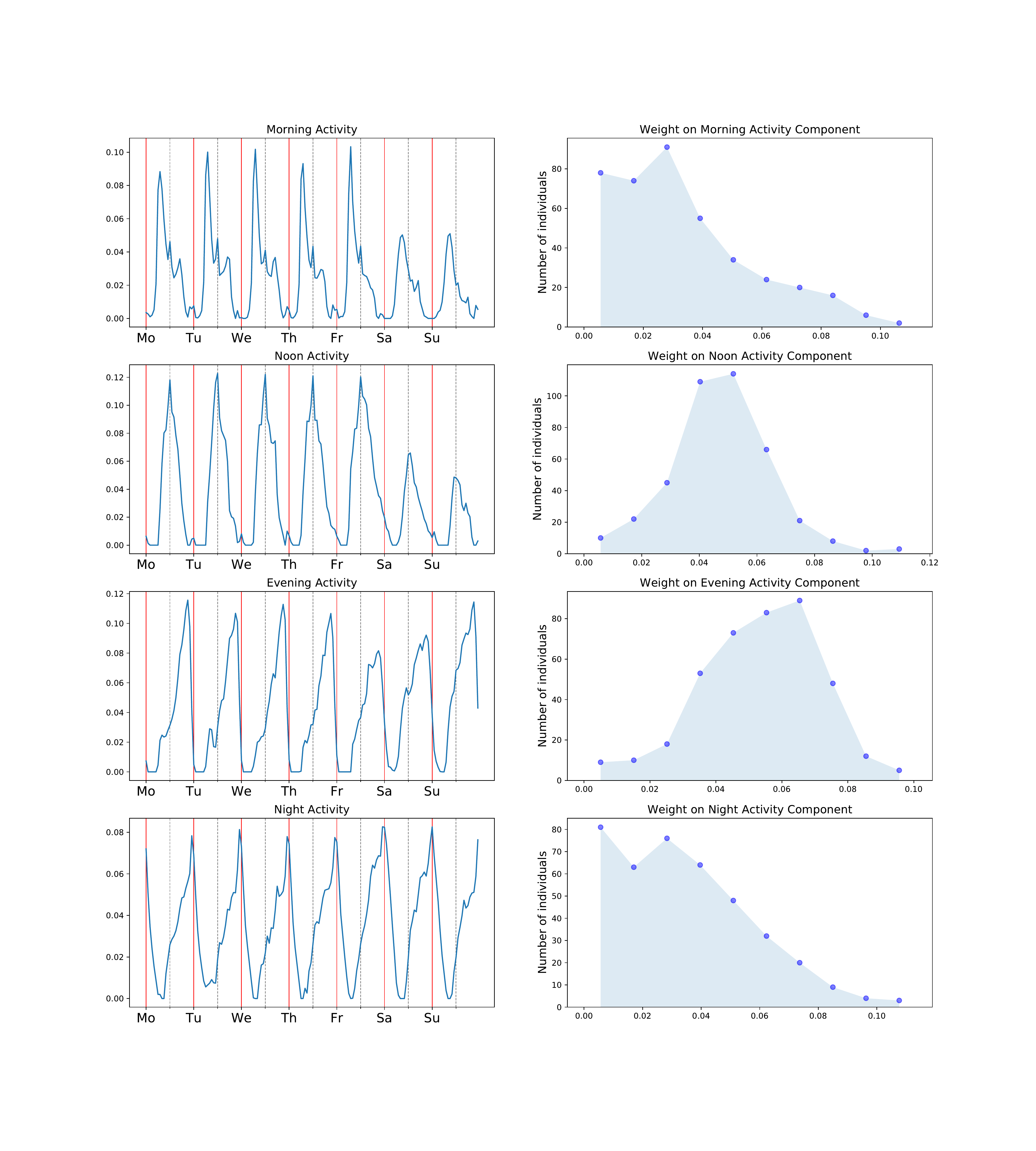}
\caption{The four NMF components (left) and the distributions of weights of individuals' activity rhythm decompositions on the four components (right). The components are detected in the data without using any external information on the daily rhythms, and we have named them after the detection as \emph{Morning activity}, \emph{Noon activity}, \emph{Evening activity}, and \emph{Night activity} for clarity. These names are chosen based on the times of the day when the activity peaks for each component.}
\label{fig:NMF-components}
\end{figure}

We used NMF as an unsupervised method to extract the four main components of the activity rhythms of the 400 students included in the analysis (for details on the choice of method and number of components see section~\ref{sec:method_choice}). 
The four NMF components are displayed in Fig.~\ref{fig:NMF-components}. 
The components follow a daily periodicity, peaking at different hours of day. 
The timings of the four peaks are associated with morning, noon, evening, and night. 

Out of the four components, the morning and noon components show lower levels of activity in the weekends compared to weekdays. 
The evening activity is lower on Friday and Saturday evenings and the night activity remains approximately similar on all days of the week. 
Taken together, as other components than nighttime activity are lower in the weekends, there is a shift in activity patterns of people towards later hours in the weekend days, which is consistent with typical behavior of adults who often go to bed earlier during the week and stay up later and catch up on sleep at the end of the week~\cite{roenneberg2003life}.

Fig.~\ref{fig:NMF-components}, right column, shows the distribution of weights on different components. 
The distribution of weights on the noon and evening components resemble  a normal distribution centered around a mean value. 
However, the distributions of the weights on the morning and night components are more skewed so that the majority of people have low weights. 

\begin{figure}[ht!]
\centering
\includegraphics[width=\linewidth]{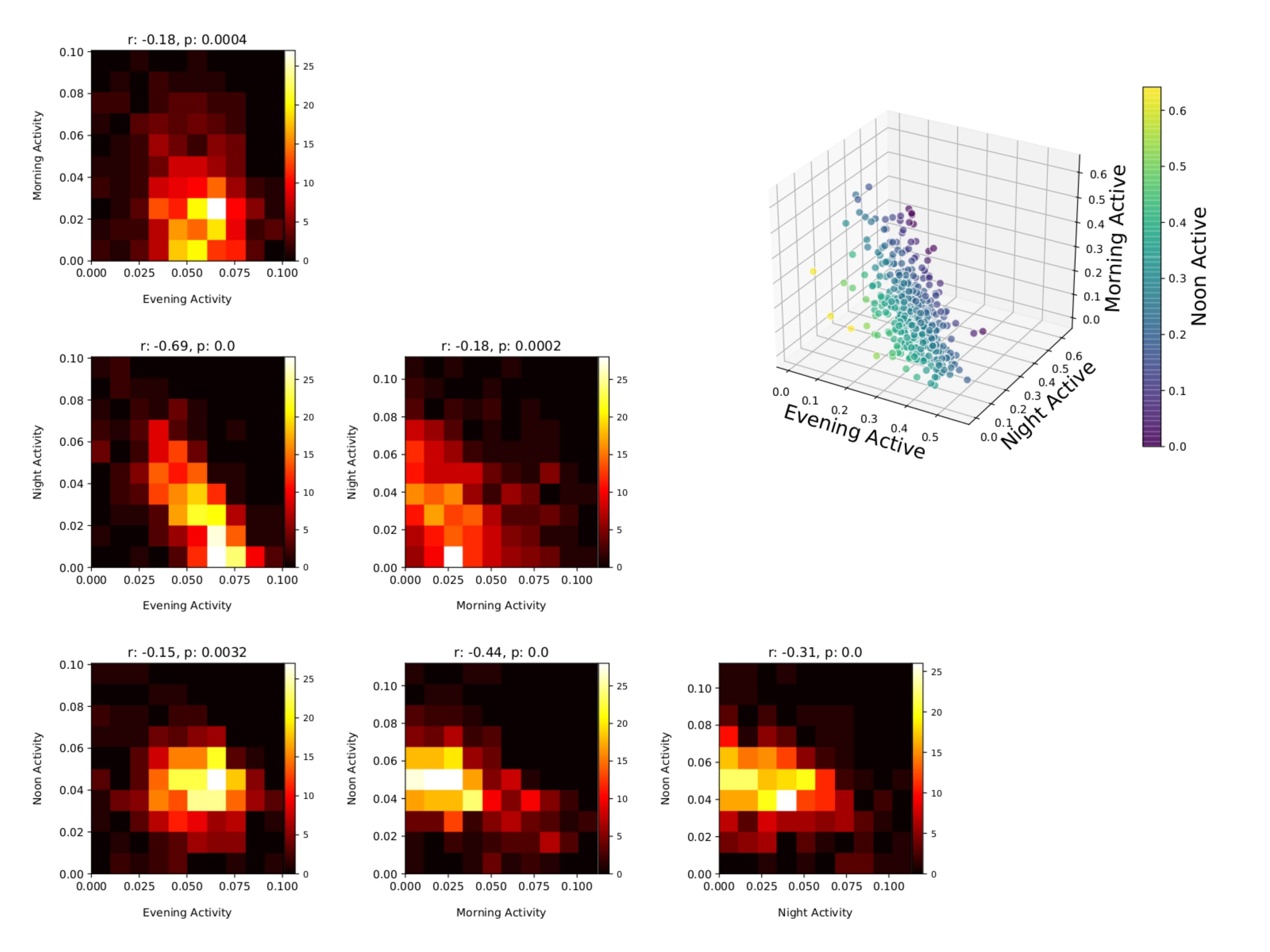}
\caption{Correlation between weight on one component vs.~the other components. The color depicts the number of people with values within each square. On top of each panel, the value of the Pearson correlation coefficient (left) and the corresponding p-value (right) are shown. The highest (anti)-correlations are between the Evening and Night components (-0.69) and Morning and Noon components (-0.44), meaning that people tend to be more active in either of each the two components for each pair. In the top right 3D plot, each dot represents one person so that the weights on three components (morning, evening  and night) are the coordinates in the 3D space and the weight on the fourth component (noon) is represented by the color of the dot. Despite the existence of some outliers, people tend to form a spectrum within this space. High values of weights for Evening activity coincide with low weights on all other dimensions. In this plot the sum of the four weights for each person is normalized to unity.}
\label{fig:comps-vs-comps}
\end{figure}

The weights in decompositions on the four NMF components are not independent. 
Fig.~\ref{fig:comps-vs-comps} explores the correlations between the weights of the decomposition of individuals' activity rhythms on the components.
While the correlation between some weights is mild (morning and evening, morning and night, noon and evening), some of the components show higher anti-correlations. 
For example, the Pearson correlation coefficients for weights on night and evening components is -0.69, which means people do not typically have high weights on both of these components simultaneously. 
Fig.~\ref{fig:comps-vs-comps}, top right, shows the weight for each person on the four components. 
In this plot we see that individuals cannot clearly be grouped into separate groups based on the activity rhythms and their weights on the four components form a continuous spectrum. 
This spectrum however has an elongated form (rather than a cloud) and shows that high weight of the evening component coincides with low values on all other components.

\begin{figure}[h!]
\centering
\includegraphics[width=\linewidth]{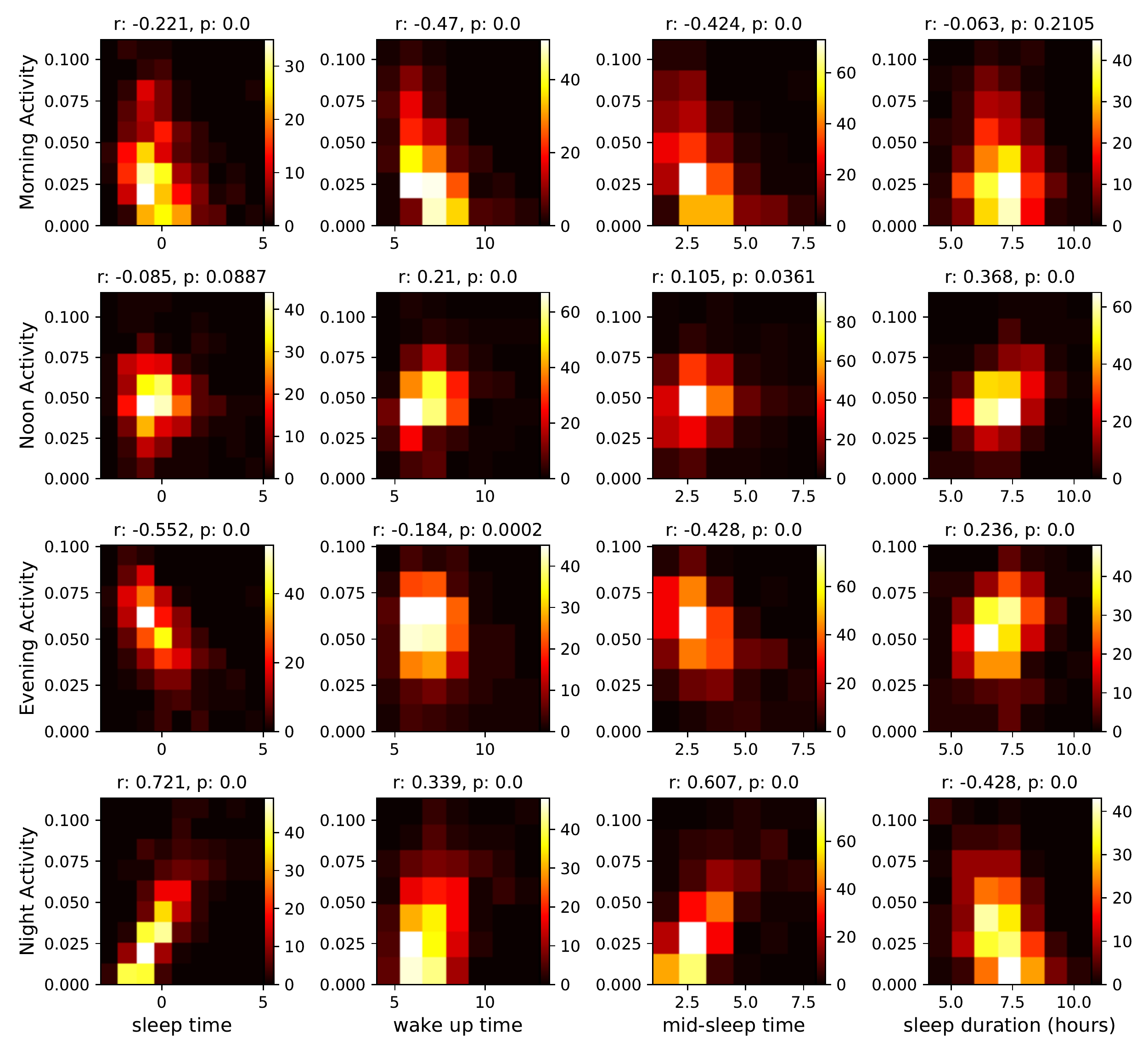}
\caption{Correlations of different NMF components with five different parameters, from right to left: sleep time, wake-up time, mid-sleep time and sleep duration.}
\label{fig:NMF-components-vs-sleep}
\end{figure}

Sleep is a substantial part of people's days and therefore one of the main determinants of how the daily rhythm of a person looks. 
Sleeping times are represented in the daily rhythms with long periods of inactivity. 
While the NMF components and the weights for different people on them are associated with their hours of activity, they are also equally associated with their hours of inactivity and sleep.
Next, we calculate the most frequent going-to-sleep time, wake-up time, mid-sleep time, and sleep duration directly from the activity data (rather than using the NMF components or the daily rhythms). 
We then examine the correlation of these variables with weights on different components (see Fig.~\ref{fig:NMF-components-vs-sleep}). 
In Section~\ref{sec:sleep-wake-times} the derivation of these variables is explained.

Sleep time shows a moderate anti-correlation with the weight on the evening activity and a high correlation with the weight on the night activity component (see table~\ref{tab:comps_sleep}). 
This implies that going to sleep later is associated with higher activity on the phone on late hours of the day. 
This finding is non-trivial because staying up later does not necessarily mean that the person has to be more active on the phone. For calculation of sleep times, even one data point at a late hour (e.g. while setting up the alarm before going to bed) would be sufficient for our algorithm to determine a late sleep time. However, for a high weight on the night activity component the person has to have many data points (screen-on activity) at late hours. This also implies that those who have a high evening activity tend to be early sleepers, while those with high night activity levels are late sleepers.
Similarly, wake-up time shows a moderate anti-correlation with the morning activity component, meaning that an earlier wake-up time is associated with higher activity level on the phone in the morning. 
The third and fourth columns in Fig.~\ref{fig:NMF-components-vs-sleep} show the mid-sleep point and  sleep duration vs weight on different components. 
The mid-sleep time is the middle point between sleep time and wake-up time, and therefore the mid-sleep time can be the same for two people with different sleep durations. 
The mid-sleep time shows moderate to high (anti-)correlations with the morning activity, evening activity and night activity components. 
Sleep duration shows a moderate anti-correlation with the night activity component and a moderate correlation with the noon activity component.

\begin{table}[ht]
\centering
\caption{Pearson correlation coefficient for different NMF components and sleep and activity parameters: sleep time, wake-up time, mid-sleep time and sleep duration.}
\begin{tabular}[t]{llcccc}
\specialrule{1.5pt}{1pt}{1pt}
& &\textbf{Sleep time}&\textbf{Wake-up time}&\textbf{Mid-sleep time}&\textbf{Sleep duration}\\
\specialrule{1.5pt}{1pt}{1pt}
&r&-0.221&-0.47&-0.424&-0.063\\
\textbf{Morning activity}&&&&&\\
&p-value&0.0&0.0&0.0&0.2105\\
\specialrule{1.5pt}{1pt}{1pt}
&r&-0.085&0.21&0.105&0.368\\
\textbf{Noon activity}&&&&&\\
&p-value&0.0887&0.0&0.0361&0.0\\
\specialrule{1.5pt}{1pt}{1pt}
&r&-0.522&-0.184&-0.428&0.236\\
\textbf{Evening activity}&&&&&\\
&p-value&0.0&0.0002&0.0&0.0\\
\specialrule{1.5pt}{1pt}{1pt}
&r&0.721&0.339&0.607&-0.428\\
\textbf{Night activity}&&&&&\\
&p-value&0.0&0.0&0.0&0.0\\

\specialrule{1.5pt}{1pt}{1pt}
\end{tabular}
\label{tab:comps_sleep}
\end{table}%
\section{Methods}\label{sec:methods}
\subsection{Dataset and Data Pre-processing}

In this work we used the same pre-processed data as in a previous study~\cite{aledavood2018social}. 
The data used were from weeks 2–51 of the year 2014. 
Data from weeks 1 and 52 were discarded, because the first week partially lies in 2013 and in week 52 the Christmas holidays lead to atypical temporal rhythms. 
There were in total $N=804$ students that used their study phones during this year. 
We excluded study participants who did not use their phone actively or did not use it at all for part of the year. 
The inclusion criteria were: (1) the person should have used the phone on $80\%$ of the days, (2) during weeks 2–51, the participant should on average have 280 screen-on and screen-off events. 
The final number of study participants which were kept for further analysis was $N=400$.

\subsection{Extracting dominant rhythms of the system with NMF}

In order to extract common temporal patterns from the data we use non-negative matrix factorization (NMF).
NMF results in a (small) number of typical patterns from a dataset such that the original data can be approximated as weighted sums of those typical patterns.
The constraint that the components must be non-negative makes interpretation of the decomposition more intuitive than e.g.~principal component analysis.
Empirically, it has been found that NMF
tends to produce components that corresponds to individual parts of a system -- when faces are decomposed, the components become eyes, noses, mouths, and moustaches~\cite{lee1999learning}.

Formally the decomposition is achieved by approximating a data matrix $\mathbf{X}$ of non-negative elements with a product of two other non-negative matrices as $\mathbf{X} \approx \mathbf{WH}^{\mathsf{T}}$, where the dimension (i.e.\ the number of columns) of the factorization matrices $\mathbf{W}$ and $\mathbf{H}$ is smaller than the dimension of the data matrix.
The dimension of the factorization means the number of typical patterns, called \emph{components}, that are sought from the data.

The approximation above is achieved by minimization of an error between the actual data matrix and the factorization.
Any meaningful error function can be used, but we used the most standard one, i.e.\ the squared Frobenius distance, which generalizes the Euclidean distance from vectors to matrices:
\begin{equation}
\label{eq:NMF_error}
E = \tfrac{1}{2}\| \mathbf{X} - \mathbf{W}\mathbf{H}^{\mathsf{T}} \|_{\mathrm{Fro}}^2
= \tfrac{1}{2}\sum_{i,j} \Big(x_{ij} - \sum_k w_{ik} h_{jk} \Big)^2.
\end{equation}

For computing NMF, we use the \texttt{scikit-learn} Python package \cite{scikit-learn}, which contains an off-the-shelf implementation of the Hierarchical Alternative Least Squares (HALS) algorithm for minimizing the error (\ref{eq:NMF_error}) \cite{cichocki2009fast}.
The implementation is stochastic, because of which we always ran the algorithm with 1000 different random seeds and picked the run that resulted in the smallest error.

In more detail, we store our activity data in an $N \times M$ matrix $\mathbf{X}$, where $N$ is the number of data vectors (in our case the number of individuals studied) and $M$ their dimensionality (the number of hours in a week).
Non-negative matrix factorization (NMF) means an approximation of $\mathbf{X}$ as
\[
\mathbf{X} \approx \mathbf{WH}^{\mathsf{T}},
\]
where $\mathbf{H}$ is an $M \times K$ and $\mathbf{W}$ an $N \times K$ matrix, and $K$ is the number of components sought from the data.
The components are the $K$ column vectors of matrix $\mathbf{H}$ (i.e.\ rows of $\mathbf{H}^{\mathsf{T}}$).
In other words, the components are simply vectors in the original data space, meaning, in our case, histograms of weekly mobile phone usage patterns.
The contribution of each component to the approximation of each data point is given by the elements of the other factor matrix, $\mathbf{W}$.
For example, the first data vector, $\mathbf{x}_1$, i.e.\ the first row of matrix $\mathbf{X}$, is approximated in NMF as
\[
\mathbf{x}_1 
= w_{11} \mathbf{h}_1^{\mathsf{T}} 
+ w_{12} \mathbf{h}_2^{\mathsf{T}}
+ \ldots 
+ w_{1K} \mathbf{h}_K^{\mathsf{T}},
\]
where $\mathbf{h}_i^{\mathsf{T}}$ are the rows of matrix $\mathbf{H}^{\mathsf{T}}$.

The complete factorization can be represented graphically as a decomposition of rank-one matrices as:
\begin{align}
\label{eq:NMF}
\begin{array}{lcc}
& \footnotesize{\color{red} M \text{ (features)}} &\\
  \footnotesize{\color{red} N \text{ (vectors)}} \hspace{-.3cm}
&\left[\begin{array}{ccccccc}
 &  &  &            &  &  &  \\
 &  &  &            &  &  &  \\ 
 &  &  & \mathbf{X} &  &  &  \\
 &  &  &            &  &  &  \\
 &  &  &            &  &  &
 \end{array}\right] 
 &\approx
 \end{array}
\begin{array}{rc}
& \footnotesize{\color{red}K\text{ (weights)}} \\
  \footnotesize{\color{red}N} \hspace{-.3cm}
&\left[\begin{array}{cl}
 &                                         \\
 &                                         \\ 
 &  \hspace{-.2cm}\mathbf{W}\hspace{.2cm}  \\
 &                                         \\
 &             
 \end{array}\right]
 \end{array}
\hspace{-.3cm}
\begin{array}{rc}
&
\footnotesize{\color{red}M} \\
\footnotesize{\color{red}K\text{  (components)}} \hspace{-.3cm}
&
\left[\begin{array}{ccccccc}
 &  &  &                         &  &  &  \\
 &  &  & \mathbf{H}^{\mathsf{T}} &  &  &  \\
 &  &  &                         &  &  &
 \end{array}\right] 
 \end{array}
 \nonumber
 \\
 \nonumber
 \\
 =
\left[\begin{array}{c}
\\ \\ 
\hspace{-.2cm}\mathbf{w}_{1}\hspace{-.2cm} \\
\\ \\ 
 \end{array}\right]
\Big[
\qquad \mathbf{h}_{1}^{\mathsf{T}} \qquad
\Big] 
+
\left[\begin{array}{c}
\\ \\ 
\hspace{-.2cm}\mathbf{w}_{2}\hspace{-.2cm} \\
\\ \\ 
 \end{array}\right]
\Big[
\qquad \mathbf{h}_{2}^{\mathsf{T}} \qquad
\Big]
+
\cdots
+
\left[\begin{array}{c}
\\ \\ 
\hspace{-.2cm}\mathbf{w}_{K}\hspace{-.2cm} \\
\\ \\ 
 \end{array}\right]
\Big[
\qquad \mathbf{h}_{K}^{\mathsf{T}} \qquad
\Big],
\end{align}
where the $i$-th value of each weight vector $\mathbf{w}_j$ indicates the contribution of component $\mathbf{h}_j$ to the representation of data vector $i$ in the reduced space.



\subsection{Choice of methods and number of components}\label{sec:method_choice}
\begin{figure}[h!]
\centering
\includegraphics[width=0.9\linewidth]{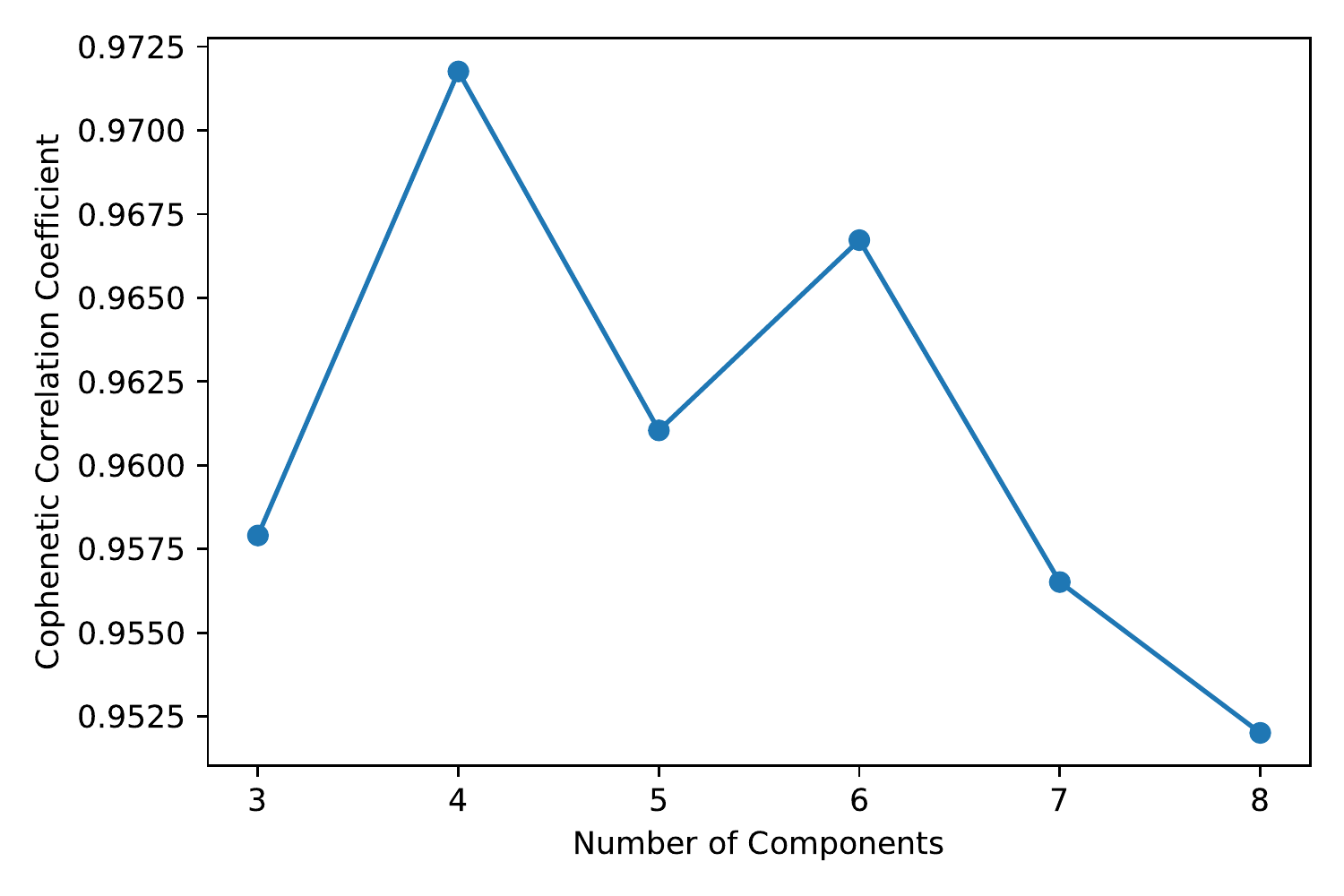}
\caption{Cophenetic correlation coefficient versus the number of components}
\label{fig:coff-corr-coef}
\end{figure}

The extraction of dominant rhythms of the system can be approached with several different tools, including principal and independent component analyses, factor analysis, topic modeling, as well as some functional data analysis methods.
We did also experiment with some of these methods, but found the results obtained with NMF most interesting and informative.
In essence, we chose NMF for its interpretability (e.g.\ compared to component analysis methods which do not have the non-negativity constraint) and its conceptual simplicity (over more involved methods such as topic modeling and functional data analysis methods).
Also, the phone screen activity data that we analyze is non-negative, which further makes NMF a natural choice. 
For the optional number of components we started with 3 components which is the number of commonly used chronotypes. 
Using the cophenetic correlation coefficient, which is a measure of finding optimal number of components in NMF based on stability of components in different runs of the algorithm, we could see that 4 is where this number is maximized (see Fig.~\ref{fig:coff-corr-coef}). 
For calculating the cophenetic correlation coefficient we used Nimfa, a Python library for NMF~\cite{Zitnik2012}.

\begin{figure}[h!]
\centering
\includegraphics[width=\linewidth]{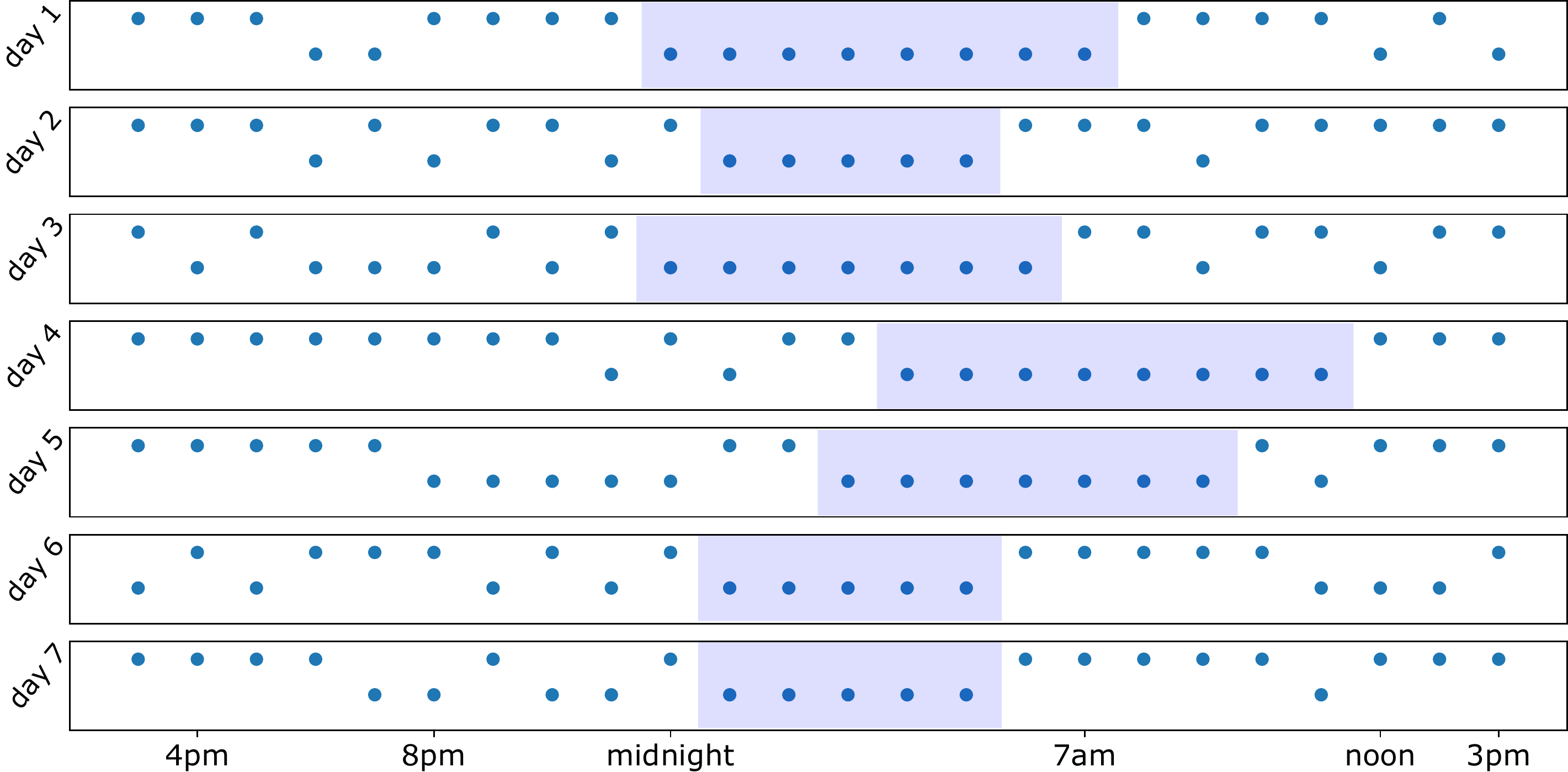}
\caption{This figure shows presences of screen-on events in different hours and different days (each row is one day of data) from one person in the study. The higher dots show presence of activity in that hour and lower dots show the lack of it. We find the longest period of of inactivity within the 24 hours (area filled with blue). These hours are marked as sleeping times. The first hour of inactivity is recognized as (going to) ``sleep time'' and the first hour right after of the period of inactivity is determined to be the ``wake-up time''. This processes repeated for all days for each person.}
\label{fig:sleep-wake}
\end{figure}

\subsection{Calculating the most common sleep and wake-up times}\label{sec:sleep-wake-times}
We assume that the longest period of inactivity on the phone during each 24-hour period for each person is their sleep time and calculate their sleep and wake-up times based on this assumption. 
This method does not give exact sleep and wake-up times for each and every night, but by looking at these values over a longer period, typical sleeping times of a person can be inferred~\cite{ciman2019smartphones, borger2019capturing}. 
In Fig.~\ref{fig:sleep-wake}, 7 days of data for one person is shown, where hours of the day when the person has been active on the phone and the hours when they have not touched the screen are depicted. 
For calculating the common sleep and wake-up times for each person, we look at data from all days, find the longest period of inactivity and pick the first hour of the longest period of inactivity as the start of the sleep time. 
Similarly we take the first hour of activity after the long period of inactivity as the wake-up time. 
For doing this, we do not go by calendar days (starting at midnight), but rather go with 24 hours starting at 3pm (assuming that sleeping is most likely happening at night hours), to reduce the chance of splitting the longest period of inactivity for the day when separating the data for consecutive days. 
We calculate the sleep, wake-up, and mid-sleep hours in this way for every day and find the most common values (mode) for each distribution. 
Using the mode we reduce the effect of outlier nights on the calculations. 
For calculating the common sleep duration for each person we simply use the number of hours of inactivity and find the mean of the distribution for each person.



\section{Discussion}

In this work we study the sleep and activity of a group of 400 university students, using data on smartphone screen-on timestamps. 
We see clear daily and weekly rhythms where there is higher levels of activity during the daytime and low activity levels at night. 
Also, we observe a delayed phase for the peak of activity and the lowest activity levels (rest times) during weekends. 
This holds both for individuals and group level activity patterns. 
Even though all patterns have a 24-hour periodicity, the times of activity peaks and lowest activity do not coincide for everybody. 
This is consistent with findings from chronobiology studies: different people have propensity to sleep at different hours of the day and the timing when they are most active or alert also varies.

We use NMF for extracting main temporal components of the data and show that four meaningful components emerge. 
The activity peak for each of these components is at a different hour of the day (morning, noon, evening, and night). 
When we decompose individuals' activity rhythms and look into the weights on the four components, we see that these weights form a continuous spectrum and do not exhibit clear-cut clusters.

Despite clear individual differences in the phase of the circadian rhythms (which are internal rhythms), exogenous factors can play a big role in how an individual behaves and when they go to sleep and wake up. 
Also, differences in cultures, climates, environment, and so on, can make a difference in who is perceived as a morning person or a night owl within a population. 
For example, waking up at 8 am in one community might result in the label `morning person' but 8 am could be considered a `later riser' within another group or population. 
While some chronotype questionnaires try to capture these external factors, they still depend on data from a pool of people to set up the cut-off thresholds for different chronotypes. 
Our method for extracting the most dominant activity rhythms in the system does not depend on arbitrary thresholds on activity hours derived from data from other populations, and it can be used for groups of almost any size. 
The analyses developed here can further be used to categorize people based on their weights on different components and give a measure of how one person compares to other people in the group under study. 
For example, we can rank people based on their weights on the morning component and label a person as a morning person within that cohort. 
However, if the available data is from too few individuals, our method would pick up  small nuances of the rhythms and would not necessarily give robust results. 
Additionally, our study suggests that even when categorizing people into groups based on their circadian patterns, it would be useful to take into account their sleep duration as well as the timing of their activity peaks during the day, in addition to their sleep timing.


A future line of study would be to look into the robustness of the components by studying different population sizes and different life styles and ages. 
It will also be informative to use a dataset where both phone activity data and chronotype questionnaires are available. 
We can compare questionnaire categories with those derived by using NMF and a clustering method on the weights on the components. 
In addition to that, we can apply the same method on other types of data gathered from phones and other wearables which have (at least) an hourly resolution. 
This allows us to study the behavioral patterns for other activity types and compare them across different data streams. 
For example, in previous work, we showed that individuals tend to have persistent daily rhythms for communication across different channels (calls and text messages) but the shapes of these rhythms and peak hours differ~\cite{aledavood2016channel}. 
Studying rhythms of different data streams we can for example investigate how rhythms of social activity vary from those of physical activity.

Some of the previous questionnaire-based works on chronotypes have pointed out that using thresholds in these questionnaires for separating different groups is rather meaningless~\cite{roenneberg2003life}, and it has been suggested that there is a wide and continuous spectrum of individual types. 
We show this based on a data-driven method and unobtrusive method rather than by using questionnaires. 
Our results do not support the use of strictly categorical variables such as chronotypes for describing people's activity and sleep habits.
Considering the chronotype as a continuous variable instead might improve the results of studies related to adverse health impacts of late phase of the circadian rhythms.  
Given the work presented here, we believe that there is a need to rethink how to quantify temporal patterns in human behavior and to extend the concept so that it takes the nuances of the activity level during non-sleep periods better into account. 

 


\bibliography{refs}

\begin{thebibliography}{10}
\urlstyle{rm}
\expandafter\ifx\csname url\endcsname\relax
  \def\url#1{\texttt{#1}}\fi
\expandafter\ifx\csname urlprefix\endcsname\relax\def\urlprefix{URL }\fi
\expandafter\ifx\csname doiprefix\endcsname\relax\def\doiprefix{DOI: }\fi
\providecommand{\bibinfo}[2]{#2}
\providecommand{\eprint}[2][]{\url{#2}}

\bibitem{foster2014rhythms}
\bibinfo{author}{Foster, R.~G.} \& \bibinfo{author}{Kreitzman, L.}
\newblock \bibinfo{journal}{\bibinfo{title}{The rhythms of life: what your body
  clock means to you!}}
\newblock {\emph{\JournalTitle{Experimental physiology}}}
  \textbf{\bibinfo{volume}{99}}, \bibinfo{pages}{599--606}
  (\bibinfo{year}{2014}).

\bibitem{panda2002circadian}
\bibinfo{author}{Panda, S.}, \bibinfo{author}{Hogenesch, J.~B.} \&
  \bibinfo{author}{Kay, S.~A.}
\newblock \bibinfo{journal}{\bibinfo{title}{Circadian rhythms from flies to
  human}}.
\newblock {\emph{\JournalTitle{Nature}}} \textbf{\bibinfo{volume}{417}},
  \bibinfo{pages}{329--335} (\bibinfo{year}{2002}).

\bibitem{edery2000circadian}
\bibinfo{author}{Edery, I.}
\newblock \bibinfo{journal}{\bibinfo{title}{Circadian rhythms in a nutshell}}.
\newblock {\emph{\JournalTitle{Physiological genomics}}}
  \textbf{\bibinfo{volume}{3}}, \bibinfo{pages}{59--74} (\bibinfo{year}{2000}).

\bibitem{irwin2015sleep}
\bibinfo{author}{Irwin, M.~R.}
\newblock \bibinfo{journal}{\bibinfo{title}{Why sleep is important for health:
  a psychoneuroimmunology perspective}}.
\newblock {\emph{\JournalTitle{Annual review of psychology}}}
  \textbf{\bibinfo{volume}{66}} (\bibinfo{year}{2015}).

\bibitem{kerkhof1985inter}
\bibinfo{author}{Kerkhof, G.~A.}
\newblock \bibinfo{journal}{\bibinfo{title}{Inter-individual differences in the
  human circadian system: a review}}.
\newblock {\emph{\JournalTitle{Biological psychology}}}
  \textbf{\bibinfo{volume}{20}}, \bibinfo{pages}{83--112}
  (\bibinfo{year}{1985}).

\bibitem{fabbian2016chronotype}
\bibinfo{author}{Fabbian, F.} \emph{et~al.}
\newblock \bibinfo{journal}{\bibinfo{title}{Chronotype, gender and general
  health}}.
\newblock {\emph{\JournalTitle{Chronobiology international}}}
  \textbf{\bibinfo{volume}{33}}, \bibinfo{pages}{863--882}
  (\bibinfo{year}{2016}).

\bibitem{antypa2016chronotype}
\bibinfo{author}{Antypa, N.}, \bibinfo{author}{Vogelzangs, N.},
  \bibinfo{author}{Meesters, Y.}, \bibinfo{author}{Schoevers, R.} \&
  \bibinfo{author}{Penninx, B.~W.}
\newblock \bibinfo{journal}{\bibinfo{title}{Chronotype associations with
  depression and anxiety disorders in a large cohort study}}.
\newblock {\emph{\JournalTitle{Depression and anxiety}}}
  \textbf{\bibinfo{volume}{33}}, \bibinfo{pages}{75--83}
  (\bibinfo{year}{2016}).

\bibitem{romo2020evening}
\bibinfo{author}{Romo-Nava, F.} \emph{et~al.}
\newblock \bibinfo{journal}{\bibinfo{title}{Evening chronotype as a discrete
  clinical subphenotype in bipolar disorder}}.
\newblock {\emph{\JournalTitle{Journal of Affective Disorders}}}
  \textbf{\bibinfo{volume}{266}}, \bibinfo{pages}{556--562}
  (\bibinfo{year}{2020}).

\bibitem{adan2012circadian}
\bibinfo{author}{Adan, A.} \emph{et~al.}
\newblock \bibinfo{journal}{\bibinfo{title}{Circadian typology: a comprehensive
  review}}.
\newblock {\emph{\JournalTitle{Chronobiology international}}}
  \textbf{\bibinfo{volume}{29}}, \bibinfo{pages}{1153--1175}
  (\bibinfo{year}{2012}).

\bibitem{levandovski2013chronotype}
\bibinfo{author}{Levandovski, R.}, \bibinfo{author}{Sasso, E.} \&
  \bibinfo{author}{Hidalgo, M.~P.}
\newblock \bibinfo{journal}{\bibinfo{title}{Chronotype: a review of the
  advances, limits and applicability of the main instruments used in the
  literature to assess human phenotype}}.
\newblock {\emph{\JournalTitle{Trends in psychiatry and psychotherapy}}}
  \textbf{\bibinfo{volume}{35}}, \bibinfo{pages}{3--11} (\bibinfo{year}{2013}).

\bibitem{horne1976self}
\bibinfo{author}{Horne, J.~A.} \& \bibinfo{author}{{\"O}stberg, O.}
\newblock \bibinfo{journal}{\bibinfo{title}{A self-assessment questionnaire to
  determine morningness-eveningness in human circadian rhythms.}}
\newblock {\emph{\JournalTitle{International journal of chronobiology}}}
  (\bibinfo{year}{1976}).

\bibitem{roenneberg2003life}
\bibinfo{author}{Roenneberg, T.}, \bibinfo{author}{Wirz-Justice, A.} \&
  \bibinfo{author}{Merrow, M.}
\newblock \bibinfo{journal}{\bibinfo{title}{Life between clocks: daily temporal
  patterns of human chronotypes}}.
\newblock {\emph{\JournalTitle{Journal of biological rhythms}}}
  \textbf{\bibinfo{volume}{18}}, \bibinfo{pages}{80--90}
  (\bibinfo{year}{2003}).

\bibitem{roenneberg2015human}
\bibinfo{author}{Roenneberg, T.} \emph{et~al.}
\newblock \bibinfo{title}{Human activity and rest in situ}.
\newblock In \emph{\bibinfo{booktitle}{Methods in enzymology}}, vol.
  \bibinfo{volume}{552}, \bibinfo{pages}{257--283}
  (\bibinfo{publisher}{Elsevier}, \bibinfo{year}{2015}).

\bibitem{cichocki2009nonnegative}
\bibinfo{author}{Cichocki, A.}, \bibinfo{author}{Zdunek, R.},
  \bibinfo{author}{Phan, A.~H.} \& \bibinfo{author}{Amari, S.-i.}
\newblock \emph{\bibinfo{title}{Nonnegative matrix and tensor factorizations:
  applications to exploratory multi-way data analysis and blind source
  separation}} (\bibinfo{publisher}{John Wiley \& Sons}, \bibinfo{year}{2009}).

\bibitem{cuttone2014inferring}
\bibinfo{author}{Cuttone, A.}, \bibinfo{author}{Lehmann, S.} \&
  \bibinfo{author}{Larsen, J.~E.}
\newblock \bibinfo{title}{Inferring human mobility from sparse low accuracy
  mobile sensing data}.
\newblock In \emph{\bibinfo{booktitle}{Proceedings of the 2014 ACM
  International Joint Conference on Pervasive and Ubiquitous Computing: Adjunct
  Publication}}, \bibinfo{pages}{995--1004} (\bibinfo{year}{2014}).

\bibitem{aledavood2015daily}
\bibinfo{author}{Aledavood, T.} \emph{et~al.}
\newblock \bibinfo{journal}{\bibinfo{title}{Daily rhythms in mobile telephone
  communication}}.
\newblock {\emph{\JournalTitle{PloS one}}} \textbf{\bibinfo{volume}{10}},
  \bibinfo{pages}{e0138098} (\bibinfo{year}{2015}).

\bibitem{aledavood2017temporal}
\bibinfo{author}{Aledavood, T.}
\newblock \bibinfo{title}{Temporal patterns of human behavior}
  (\bibinfo{year}{2017}).

\bibitem{urena2020going}
\bibinfo{author}{Ure{\~n}a-Carrion, J.}, \bibinfo{author}{Saram{\"a}ki, J.} \&
  \bibinfo{author}{Kivel{\"a}, M.}
\newblock \bibinfo{journal}{\bibinfo{title}{Going beyond communication
  intensity for estimating tie strengths in social networks}}.
\newblock {\emph{\JournalTitle{arXiv:2007.14238}}}  (\bibinfo{year}{2020}).

\bibitem{stopczynski2014measuring}
\bibinfo{author}{Stopczynski, A.} \emph{et~al.}
\newblock \bibinfo{journal}{\bibinfo{title}{Measuring large-scale social
  networks with high resolution}}.
\newblock {\emph{\JournalTitle{PloS one}}} \textbf{\bibinfo{volume}{9}},
  \bibinfo{pages}{e95978} (\bibinfo{year}{2014}).

\bibitem{eagle2006reality}
\bibinfo{author}{Eagle, N.} \& \bibinfo{author}{Pentland, A.~S.}
\newblock \bibinfo{journal}{\bibinfo{title}{Reality mining: sensing complex
  social systems}}.
\newblock {\emph{\JournalTitle{Personal and ubiquitous computing}}}
  \textbf{\bibinfo{volume}{10}}, \bibinfo{pages}{255--268}
  (\bibinfo{year}{2006}).

\bibitem{wang2014studentlife}
\bibinfo{author}{Wang, R.} \emph{et~al.}
\newblock \bibinfo{title}{Studentlife: assessing mental health, academic
  performance and behavioral trends of college students using smartphones}.
\newblock In \emph{\bibinfo{booktitle}{Proceedings of the 2014 ACM
  international joint conference on pervasive and ubiquitous computing}},
  \bibinfo{pages}{3--14} (\bibinfo{year}{2014}).

\bibitem{mattingly2019tesserae}
\bibinfo{author}{Mattingly, S.~M.} \emph{et~al.}
\newblock \bibinfo{title}{The tesserae project: Large-scale, longitudinal, in
  situ, multimodal sensing of information workers}.
\newblock In \emph{\bibinfo{booktitle}{Extended Abstracts of the 2019 CHI
  Conference on Human Factors in Computing Systems}}, \bibinfo{pages}{1--8}
  (\bibinfo{year}{2019}).

\bibitem{monsivais2017seasonal}
\bibinfo{author}{Monsivais, D.}, \bibinfo{author}{Bhattacharya, K.},
  \bibinfo{author}{Ghosh, A.}, \bibinfo{author}{Dunbar, R.~I.} \&
  \bibinfo{author}{Kaski, K.}
\newblock \bibinfo{journal}{\bibinfo{title}{Seasonal and geographical impact on
  human resting periods}}.
\newblock {\emph{\JournalTitle{Scientific reports}}}
  \textbf{\bibinfo{volume}{7}}, \bibinfo{pages}{1--10} (\bibinfo{year}{2017}).

\bibitem{cuttone2017sensiblesleep}
\bibinfo{author}{Cuttone, A.} \emph{et~al.}
\newblock \bibinfo{journal}{\bibinfo{title}{Sensiblesleep: A bayesian model for
  learning sleep patterns from smartphone events}}.
\newblock {\emph{\JournalTitle{PloS one}}} \textbf{\bibinfo{volume}{12}},
  \bibinfo{pages}{e0169901} (\bibinfo{year}{2017}).

\bibitem{aledavood2018social}
\bibinfo{author}{Aledavood, T.}, \bibinfo{author}{Lehmann, S.} \&
  \bibinfo{author}{Saram{\"a}ki, J.}
\newblock \bibinfo{journal}{\bibinfo{title}{Social network differences of
  chronotypes identified from mobile phone data}}.
\newblock {\emph{\JournalTitle{EPJ Data Science}}}
  \textbf{\bibinfo{volume}{7}}, \bibinfo{pages}{1--13} (\bibinfo{year}{2018}).

\bibitem{aledavood2019smartphone}
\bibinfo{author}{Aledavood, T.} \emph{et~al.}
\newblock \bibinfo{journal}{\bibinfo{title}{Smartphone-based tracking of sleep
  in depression, anxiety, and psychotic disorders}}.
\newblock {\emph{\JournalTitle{Current psychiatry reports}}}
  \textbf{\bibinfo{volume}{21}}, \bibinfo{pages}{49} (\bibinfo{year}{2019}).

\bibitem{martinez2020improved}
\bibinfo{author}{Martinez, G.~J.} \emph{et~al.}
\newblock \bibinfo{title}{Improved sleep detection through the fusion of phone
  agent and wearable data streams}.
\newblock In \emph{\bibinfo{booktitle}{2020 IEEE International Conference on
  Pervasive Computing and Communications Workshops (PerCom Workshops)}},
  \bibinfo{pages}{1--6} (\bibinfo{organization}{IEEE}, \bibinfo{year}{2020}).

\bibitem{lee1999learning}
\bibinfo{author}{Lee, D.~D.} \& \bibinfo{author}{Seung, H.~S.}
\newblock \bibinfo{journal}{\bibinfo{title}{Learning the parts of objects by
  non-negative matrix factorization}}.
\newblock {\emph{\JournalTitle{Nature}}} \textbf{\bibinfo{volume}{401}},
  \bibinfo{pages}{788--791} (\bibinfo{year}{1999}).

\bibitem{scikit-learn}
\bibinfo{author}{Pedregosa, F.} \emph{et~al.}
\newblock \bibinfo{journal}{\bibinfo{title}{Scikit-learn: Machine learning in
  {P}ython}}.
\newblock {\emph{\JournalTitle{Journal of Machine Learning Research}}}
  \textbf{\bibinfo{volume}{12}}, \bibinfo{pages}{2825--2830}
  (\bibinfo{year}{2011}).

\bibitem{cichocki2009fast}
\bibinfo{author}{Cichocki, A.} \& \bibinfo{author}{Phan, A.-H.}
\newblock \bibinfo{journal}{\bibinfo{title}{Fast local algorithms for large
  scale nonnegative matrix and tensor factorizations}}.
\newblock {\emph{\JournalTitle{IEICE transactions on fundamentals of
  electronics, communications and computer sciences}}}
  \textbf{\bibinfo{volume}{92}}, \bibinfo{pages}{708--721}
  (\bibinfo{year}{2009}).

\bibitem{Zitnik2012}
\bibinfo{author}{Zitnik, M.} \& \bibinfo{author}{Zupan, B.}
\newblock \bibinfo{journal}{\bibinfo{title}{Nimfa: A python library for
  nonnegative matrix factorization}}.
\newblock {\emph{\JournalTitle{Journal of Machine Learning Research}}}
  \textbf{\bibinfo{volume}{13}}, \bibinfo{pages}{849--853}
  (\bibinfo{year}{2012}).

\bibitem{ciman2019smartphones}
\bibinfo{author}{Ciman, M.} \& \bibinfo{author}{Wac, K.}
\newblock \bibinfo{journal}{\bibinfo{title}{Smartphones as sleep duration
  sensors: Validation of the isensesleep algorithm}}.
\newblock {\emph{\JournalTitle{JMIR mHealth and uHealth}}}
  \textbf{\bibinfo{volume}{7}}, \bibinfo{pages}{e11930} (\bibinfo{year}{2019}).

\bibitem{borger2019capturing}
\bibinfo{author}{Borger, J.~N.}, \bibinfo{author}{Huber, R.} \&
  \bibinfo{author}{Ghosh, A.}
\newblock \bibinfo{journal}{\bibinfo{title}{Capturing sleep--wake cycles by
  using day-to-day smartphone touchscreen interactions}}.
\newblock {\emph{\JournalTitle{NPJ digital medicine}}}
  \textbf{\bibinfo{volume}{2}}, \bibinfo{pages}{1--8} (\bibinfo{year}{2019}).

\bibitem{aledavood2016channel}
\bibinfo{author}{Aledavood, T.} \emph{et~al.}
\newblock \bibinfo{title}{Channel-specific daily patterns in mobile phone
  communication}.
\newblock In \emph{\bibinfo{booktitle}{Proceedings of ECCS 2014}},
  \bibinfo{pages}{209--218} (\bibinfo{publisher}{Springer},
  \bibinfo{year}{2016}).

\end{thebibliography}



\section*{Acknowledgements}

TA and JS acknowledge support from the Academy of Finland, project number 297195. TA also acknowledges support from James S. McDonnell Foundation and thanks Mikko Kivel\"a for providing constructive feedback on the manuscript. 


\section*{Author contributions statement}
SL collected the data. TA, SJ and JS designed the study. TA, SL, IK, JS conducted the study. TA analyzed the data. All authors contributed to the writing of the manuscript. 
\section*{Additional information}

\textbf{Competing interests} The authors report no competing interests.

\end{document}